\documentclass[twocolumn,pra,showpacs,floatfix]{revtex4}
\usepackage{amsmath}
\usepackage{amsfonts}
\usepackage{amsthm}
\usepackage{graphicx}
\newcommand{\ket}[1]{\ensuremath{|#1\rangle}}
\newcommand{\bra}[1]{\ensuremath{\langle #1|}}

\newcommand{\pHH}{\textit{para}-hydrogen}

\newcommand{\diagonal}[1]{\,\text{Dg}[#1]}
\newcommand{\bdiagonal}[1]{\left\{\left\{#1\right\}\right\}}
\newcommand{\bdiagonall}[1]{\Biggl\{\Biggl\{#1\Biggr.\Biggr.}
\newcommand{\bdiagonalr}[1]{\Biggl.\Biggl.#1\Biggr\}\Biggr\}}
\newcommand{\terms}[2]{\ensuremath{\stackrel{#1\ {\rm {terms}}}{\overbrace{#2}}}}

\begin{document}
\title{Sharing polarization within quantum subspaces}
\author{M.~S.~Anwar}\email{muhammad.anwar@physics.ox.ac.uk}
\affiliation{Centre for Quantum Computation, Clarendon Laboratory,
University of Oxford, Parks Road, OX1 3PU, United Kingdom}
\affiliation{National Centre for Physics, Quaid-e-Azam University,
Islamabad, Pakistan}
\author{J.~A.~Jones}\email{jonathan.jones@qubit.org}
\affiliation{Centre for Quantum Computation, Clarendon Laboratory,
University of Oxford, Parks Road, OX1 3PU, United Kingdom}
\author{S.~B.~Duckett}\email{sbd3@york.ac.uk}
\affiliation{Department of Chemistry, University of York, Heslington, York, YO10 5DD, United Kingdom}
\date{\today}
\begin{abstract}
Given an ensemble of $n$ spins, at least some of which are partially
polarized, we investigate the sharing of this polarization within a
subspace of $k$ spins. We assume that the sharing results in a
pseudopure state, characterized by a single purity parameter which
we call the bias. As a concrete example we consider ensembles of
spin-$1/2$ nuclei in liquid-state nuclear magnetic resonance (NMR)
systems. The shared bias levels are compared with some current
entanglement bounds to determine whether the reduced subspaces can
give rise to entangled states.
\end{abstract}
\maketitle

\section{Introduction}
Liquid-state nuclear magnetic resonance (NMR) techniques have proved
a convenient test-bed for implementing quantum algorithms
\cite{cory96,cory97,jones98a,chuang98a,jones01a,vandersypen01}.
However, a conventional NMR quantum computer starts off in an
initial state which is very similar to a maximally mixed state, due
to the tiny population differences between the energy levels of the
spin-$1/2$ nuclei which act as the qubits \cite{jones00,warren97}.

Various approaches to this initialization problem have been
suggested. Firstly, it is possible to convert thermal states into
pseudopure states \cite{cory96,cory97}: this approach has been
adopted in the vast majority of NMR implementations of QIP, but is
not scalable \cite{warren97}. Secondly, non-Boltzmann distributions
can be prepared, for example with \pHH\
\cite{anwar04,anwarDeutsch,anwarGrover}, giving almost pure states
which lie above the entanglement threshold \cite{peres96}. Thirdly,
computational schemes have been described
\cite{schulman99,boykin02,fernandez04} which concentrate the small
polarization available on a large number of spins into a smaller
subspace.

Here we investigate the effects of polarization sharing on achieving
states of useful purity. The schemes mentioned above
\textit{concentrate} polarization onto a \textit{smaller} subset of
spins, but we study the effects of spreading the polarization over a
\textit{larger} subset of spins. We assume that the polarization
sharing results in a pseudopure state with a purity characterized by
a bias parameter $\delta$ that we define below. The bias is then
compared with known entanglement bounds derived for pseudopure
states.

\section{Pseudopure states and entanglement}\label{section-ps}
In the high temperature approximation, the initial state of a
(homonuclear) NMR quantum computer is
\begin{equation}\label{rho-eq-thermal}
\rho_{eq}\approx\frac{1}{2^n}\left(\mathbf{1}_{n}+\frac{B}{2}\sum_{j=1}^n\sigma_{iz}\right),
\end{equation}
where $n$ is the total number of spins, $\mathbf{1}_n$ is the
identity matrix of order $2^n$, $B=-\hbar \omega/k_BT$ is a
Boltzmann factor and $\sigma_{jz}$ is the $z$ Pauli matrix for spin
$j$. A standard implementation of an algorithm requires the quantum
computer to start off in a pure state, characterised by having a
single non-zero eigenvalue, of size one. Clearly, the thermal state
\eqref{rho-eq-thermal} has many different eigenvalues and so to
prepare pure from thermal states, a non-unitary process must be
employed at some stage. Cooling to the ground state is an example of
such a process, but is ineffective in NMR due to the small energy
gaps involved. Pseudopure $k$-qubit states have an eigenvalue
spectrum between these extremes, having one large eigenvalue and
$2^k-1$ degenerate smaller eigenvalues.  This means that thermal
states can be converted into pseudopure states by particularly
simple non-unitary processes, such as averaging $2^k-1$ population
terms.

A general $k$-qubit pseudopure state $\chi$ takes the form
\begin{equation}\label{rho-pseudo}
\chi_{k,\delta}=(1-\delta)\frac{\mathbf{1}_k}{2^k}+\delta
\ket{\psi}\bra{\psi},
\end{equation}
with dynamics identical to those of the corresponding pure state
$\ket{\psi}\bra{\psi}$. (We explicitly use the symbol $\chi$ to
denote pseudopure states.) The state is characterised by a single
parameter, $\delta$, which we call the bias of the state. Another
useful description is to write out the pseudopure state explicitly
in its eigenbasis with the eigenvalues in descending order to give
\begin{equation}\label{rho-pseudo-explicit}
\chi_{k,\delta}=\diagonal{f,\stackrel{(2^k-1)\ {\rm
{terms}}}{\overbrace{\frac{1-f}{2^k-1},\frac{1-f}{2^k-1},\ldots
,\frac{1-f}{2^k-1}}}}
\end{equation}
where $\text{Dg}[\ldots]$ denotes a diagonal matrix and $f$ is both
the fractional population of the desired state
$\ket{\psi}\bra{\psi}$ and the \textit{largest} eigenvalue and is
related to the bias by
\begin{equation}\label{delta-fvalue}
\delta=\frac{2^kf-1}{2^k-1}.
\end{equation}

A pseudopure state will always be separable if $\delta$ is less than
some critical value, and Braunstein \textit{et~al.}
\cite{braunstein99} have described upper and lower bounds on
separability. A state of the form given in Eq.~\ref{rho-pseudo} was
shown to be explicitly separable for sufficiently small biases such
that
\begin{equation}\label{lower-bound-e}
\delta \leq  \delta_l=\frac{1}{1+2^{2k-1}}
\end{equation}
and we say that these states belong to the region \textbf{S}.
If the bias exceeds their upper bound
\begin{equation}\label{upper-bound-e}
\delta > \delta_u=\frac{1}{1+2^{\frac{k}{2}}}
\end{equation}
the state is said to lie in the entangled region \textbf{E}, or more
appropriately the \textit{entanglable} \cite{yu04} region. In the
region in between, \textbf{ES}, it is not known whether entangled
states can be prepared. These bounds were subsequently improved,
shrinking the size of \textbf{ES}. For example, Gurvits and Barnum
\cite{gurvits04} have tightened the lower bound to
\begin{equation}\label{lower-bound-e-gb}
\delta_l=\,\frac{3}{2(6)^{k/2}}.
\end{equation}
Current NMR implementations with Boltzmann initialization use states
lying in the separable region \textbf{S} and it seems possible to
enter the region \textbf{ES} only by employing more qubits. For
example, using the Gurvits--Barnum lower bound with a typical
Boltzmann factor of $B=10^{-5}$, the state will cross over into
\textbf{ES} for $k\ge 41$.

Similar themes have been taken up by other authors as well.  Yu
\textit{et al.} have shown \cite{yu04} that by using unitarily
transformed thermal states in place of pseudopure states, the
\textbf{ES} to \textbf{E} transition can take place with a smaller
number of qubits: making pseudopure states from thermal states
involves convex mixing and, therefore, decreases the likelihood of
entanglement. Other researchers have shown \cite{bose02} that
entanglement can also exist in a $2\times N$ dimensional quantum
system, when \textit{only} the qubit is pure and the $N$ dimensional
system is in a highly mixed state, such as a two level atom
interacting with a high temperature field.

Here we investigate the \textit{sharing} of polarization within a
pseudopure quantum subspace.
We consider an initial state
\begin{equation}\label{rho-start}
\rho_{n,p,\sigma } = \left( \bigotimes_{j=1}^p \phi_{\sigma}\right)
\otimes \left( \bigotimes_{j=p+1}^n \mathbf{1}_1/2 \right)
\end{equation}
where $n$ is the total number of qubits of which $p$ qubits are in
the state,
\begin{equation}\label{pure}
\phi_{\sigma}= \left( \begin{array}{cc} (1+\sigma)/2&0\\
0& (1-\sigma)/2\end{array}\right),\quad 0 \leq \sigma \leq 1,
\end{equation}
having a polarization $\sigma$ and the remaining $n-p$ qubits are
maximally mixed. Our goal is to compute the achievable bias in a $k$
qubit pseudopure state where $k\le n$, \textit{i.e.}, we are
interested in the transformation
$\rho_{n,p,\sigma}\mapsto\chi_{k,\delta}$. We achieve this in two
steps: (a) convert $\rho_{n,p,\sigma}$ into a $k$ qubit state
$\tau_k$ using a partial trace operation, and then (b) convert
$\tau_k$ into a pseudopure state $\chi_{k,\delta}$ using cyclic
averaging.

For simplicity consider a state $\rho_{n,p,1}$ with all the $p$
qubits being perfectly polarized,
\begin{equation}\label{rho-target}
\begin{split}
\rho_{n,p,1} &= \left( \bigotimes_{j=1}^p \ket{0}\bra{0}\right)
\otimes \left( \bigotimes_{j=p+1}^n \mathbf{1}_1/2 \right)\\
&=\frac{1}{2^{n-p}}\diagonal{\stackrel{2^{n-p}\ {\rm
{terms}}}{\overbrace{{1,1,\ldots ,1}}},0,0,\ldots ,0}.
\end{split}
\end{equation}
and consider the case $k=p$. Partially tracing \cite{nielsen00} out
the $n-k$ qubits from $\rho_{n,p,1}$ leaves us with the $k$ qubit
reduced state
\begin{equation}
\tau_k=\diagonal{1,\stackrel{2^{k}-1\ {\rm
{terms}}}{\overbrace{{0,0,\ldots ,0}}}}.
\end{equation}
This is also a pseudopure state $\chi_{k,1}$ with maximum achievable
bias of one. The partial trace operation reduces the dimensionality
of the system and for a diagonal state, is equivalent to taking
partial sums of consecutive eigenvalues along the ordered diagonal.
Physically, it just corresponds to ignoring the $n-k$ qubits. In the
case of $k=p$, the reduced subspace after the partial trace
operation is already a (pseudo)pure state, with $\delta=1$. This
will also be the case when $k<p$. However, for $p<k\le n$, the
reduced state $\tau_k$ will have several different eigenvalues.

We can explicitly generate the desired pseudopure state by writing
the mixed state in its ordered eigenbasis and then averaging over
cyclic permutations of the $2^k-1$ trailing elements.  This
procedure (which can be considered as a generalisation of the twirl
operation \cite{anwar05}) is experimentally implementable by
exhaustive temporal averaging \cite{knill98}; more efficient
procedures are also available \cite{knill98,knill00}. A non-unitary
process cannot increase the maximum eigenvalue of a state $\tau$,
but by definition, a pseudopure state has only one non-degenerate
eigenvalue which must coincide with the maximum eigenvalue of
$\tau$. Cyclic averaging leaves this eigenvalue unchanged and
therefore extracts the maximum bias $\delta$.

\begin{figure}
\begin{center}
\includegraphics[scale=0.5]{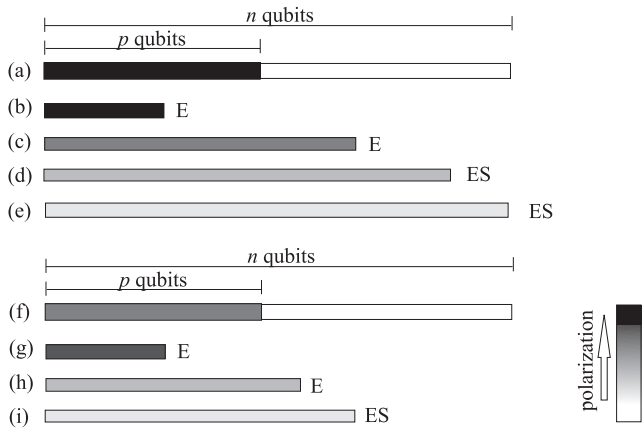}
\caption{The effect of sharing the polarization of $p$ pure
[(a)--(e)] or partially pure [(f)--(i)] spins on quantum subspaces.
Starting with $p$ pure spins (a), we can \textit{pick} $k\le p$ pure
spins (b) and extract all the available polarization. The
polarization can also be \textit{spread} onto a bigger subspace
while retaining the entanglability (c). Sharing the polarization
within even higher dimensional subspaces decreases the resulting
bias and therefore the possibility of entanglement, (d) and (e). As
the size of the extracted subspace increases, the bias drops
exponentially. Smaller biases are shown by lighter tones of grey.
For $p$ partially polarized spins (f), we can \textit{concentrate}
the bias onto smaller subspaces (g) or spread it onto bigger
subspaces, (h) and (i), with a corresponding decrease in the bias
reducing the likelihood of entanglement. \label{graphics1}}
\end{center}
\end{figure}

Algebraic manipulation shows that the maximum eigenvalue of
$\chi_{k,\delta}$ is given by the formula
\begin{equation}\label{fidelity-def}
f=\frac{\text{Tr}(\rho_{n,p,\sigma}\rho_{n,k,1})}{\text{Tr}(\rho_{n,k,1}^2)}=\frac{1}{2^{n-k}}\text{Tr}(\rho_{n,p,\sigma}\rho_{n,k,1})
\end{equation}
which is the overlap between the states (\ref{rho-start}) and
(\ref{rho-target}). Note that we do not consider here precisely
\textit{how} the polarization sharing procedures might be
implemented in practice, but simply determine the limits. Our
results are summarised in Fig.~\ref{graphics1} and exemplified
below.

\section{Perfectly polarized
qubits}\label{section-share-pure} Considering the form of Equation
\ref{rho-target} ($n$ spins out of which $p$ are perfectly polarized
and the rest are maximally mixed), we see that $f$ is given by
\begin{equation}\label{f-ptok}
f=
\begin{cases}
2^{p-k} & {\text {for}}\ k\ge p\\
1       & {\text {for}}\ k<p
\end{cases}
\end{equation}
and from Equation~\ref{delta-fvalue} the achievable bias is
\begin{equation}\label{delta-ptok}
\delta=
\begin{cases}
\frac{2^{p}-1}{2^k-1} & {\text {for}}\ k\ge p\\
1                     & {\text {for}}\ k<p.
\end{cases}
\end{equation}
For $k<p$, \textit{all} the purity can be extracted as this is
equivalent to picking out the pure spins from a set of pure and
maximally mixed spins. The case $k>p$ is more interesting as this
involves distributing the polarization of $p$ spins over a larger
subspace and is the polarization sharing that we shall be mainly
interested in.
\begin{figure}
\begin{center}
\includegraphics[scale=0.4]{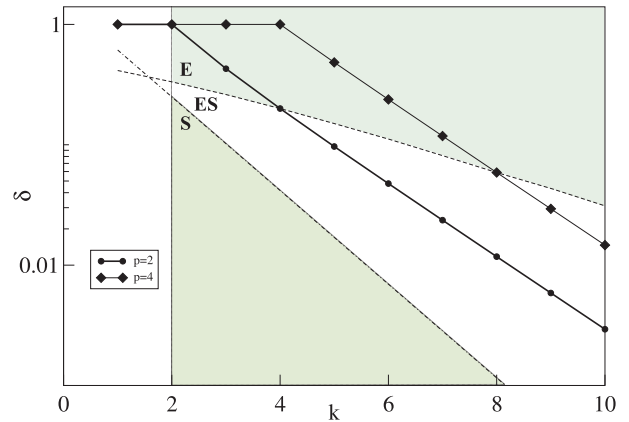}
\caption{(Color online) Sharing the polarization of $2$ and $4$ pure
spins. The upper shaded region is \textbf{E} and the lower shaded
region is \textbf{S}, whereas \textbf{ES} is in between. The border
between two regions belongs to the less entangled region. The
regions are intended to guide the eye, and are only shown for $2$ or
more spins, where the concept of entanglement is
valid.\label{graphics2}}
\end{center}
\end{figure}
Figure \ref{graphics2} shows the results for $p$ pure spins, with
$p=2$ and $4$. For $k>p$, the bias falls off exponentially as $k$ is
increased.

Examining Equation~\ref{rho-target} also shows that adding extra
qubits (increasing $n$) neither hinders nor helps the extraction of
purity onto qubits; it seems that the only role of $n$ is that it
limits the number of spins in the pseudopure subspace, as evidently
$k\le n$. Of course these additional qubits may assist the
implementation of the manipulations necessary to prepare the
pseudopure state, but they have no ultimate effect on the extracted
bias.

The biases of the $k$-qubit pseudopure subspaces can now be compared
with the entanglement bounds, but for the possibility of
entanglement to arise in the first place, the pure component in the
pseudopure state must be entangled. For example, the direct product
$2$-qubit state $\ket{00}\bra{00}$ is pure, but is clearly
separable. However, one can always find a unitary operator that
converts an arbitrary pure state into any desired pure target state,
be it entangled or otherwise. Any pure state is therefore
\textit{unitarily equivalent} to an entangled state of the same
dimensions, and we can assume that our $k$-qubit pseudopure state
comprises a pure, entangled component mixed in with the maximally
mixed state. This is the justification for using the term
\textit{entanglable} for the region \textbf{E}, as mentioned in the
previous Section.

For $k\le p$, the $k$-qubit subspace is obviously in the region
\textbf{E} but for $k>p$, increasing $k$ leads to an exponential
drop in the extractable bias $\delta$. At some critical $k$ we step
into the region \textbf{ES}. This transition from \textbf{E} to
\textbf{ES} takes place where $\delta=\delta_u$, which can be
identified in Figure \ref{graphics2} as the point at which the bias
curve crosses into the region \textbf{ES}, and occurs at $k_c=2p$.
What this means is that if we start with $p$ perfectly polarized
spins, then for $k<2p$ our extracted system will be in \textbf{E}.
However, for higher values of $k$ we shall be in \textbf{ES}.  We
also note that as a result of this polarization sharing we will
never enter \textbf{S}; this is clear from the slopes of the lines
in the figures, and a simple proof is given in the Appendix.

\section{Partially polarized
qubits}\label{section-share-impure} We now generalize the above
example to consider $p$ spins that are not perfectly polarized, each
having a uniform non-zero polarization $0<\sigma<1$. The $n$-qubit
state $\rho_{n,p,\sigma}$ is now given by
\begin{equation}\label{rho-nksigma}
\begin{split}
\rho_{n,p,\sigma}&=\left(
{\bigotimes_{j=1}^p}\bdiagonal{\frac{1+\sigma}{2},\frac{1-\sigma}{2}}\right)
\otimes\left({\bigotimes_{j=p+1}^n}\frac{\mathbf{1}_1}{2}\right)\\
&=\frac{1}{2^{n-p}}\bdiagonall{\terms{2^{n-p}}{\left(\frac{1+\sigma}{2}\right)^p},
\terms{(2^{n-p}\
^pC_1)}{\left(\frac{1+\sigma}{2}\right)^{p-1}\left(\frac{1-\sigma}{2}\right)},\\
&\ldots}, \bdiagonalr{\terms{(2^{n-p}\
^pC_{p-1})}{\left(\frac{1+\sigma}{2}\right)\left(\frac{1-\sigma}{2}\right)^{p-1}},\terms{2^{n-p}}{\left(\frac{1-\sigma}{2}\right)^p}},
\end{split}
\end{equation}
where $^lC_j=l!/j!(l-j)!$ is the binomial coefficient. If $k\geq p$,
calculating $f$ involves summing only the first $2^{n-k}$ terms,
each being of the same size $((1+\sigma)/2)^p$.  This is best
illustrated with a numerical example. Consider a state with $n=4$
and $p=2$
\begin{equation}\label{rho-42sigma}
\rho_{4,2,\sigma}
=\frac{1}{4}\diagonal{\terms{4}{\left(\frac{1+\sigma}{2}\right)^2},
\terms{8}{\left(\frac{1-\sigma}{2}\right)\left(\frac{1+\sigma}{2}\right)},\terms{4}{\left(\frac{1-\sigma}{2}\right)^2}
}
\end{equation}
and suppose we want to share the purity over a $3$-qubit subspace
($k=3$).  Calculating $f$ will only involve a partial sum of the
first $2^{n-k}=2$ terms in (\ref{rho-42sigma}), each of these terms
being $(1/4)((1+\sigma)/2)^2$. Similarly if $k=2$, we need a partial
sum over the first $4$ terms and for $k=4$, we need to consider just
the first term. In each of these cases only the leading terms are
involved in the partial sums, and we can derive a formula for $f$
\begin{equation}\label{fidelity-kgreaterp-sigma}
f=2^{-k}(1+\sigma)^p
\end{equation}
with a corresponding bias
\begin{equation}
\delta=\frac{(1+\sigma)^p-1}{2^k-1}\label{delta-kgreaterp-sigma}.
\end{equation}
If $k<p$ we are concentrating polarization onto a subspace smaller
than the original, and can expect to extract a higher bias.  in this
case the overlap and bias must be calculated explicitly on a case by
case basis. We do not consider this polarization concentration
further.

\begin{figure}
\begin{center}
\includegraphics[scale=0.4]{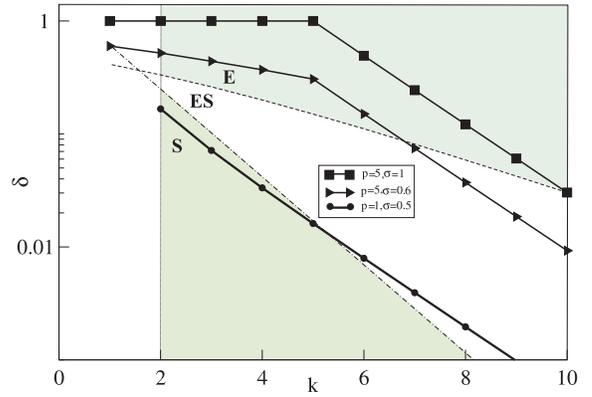}
\caption{(Color online) Sharing the polarization of impure spins.
The regions \textbf{E}, \textbf{ES} and \textbf{S} are highlighted
in the same way as in Fig. \ref{graphics2}. The figure demonstrates
polarization sharing for $5$ pure and $5$ impure spins. The effect
of sharing from a single weakly polarized spin is also shown; when
shared over $6$ or more spins the state moves from being provably
separable into the region where it is unknown whether or not it may
be entangled.Note, however, that states currently placed in
\textbf{ES} may subsequently move to \textbf{S} if lower bounds are
further tightened.\label{graphics3}}
\end{center}
\end{figure}

As with perfectly polarized spins, sharing from impure spins also
entails an exponential decrease in the extractable bias as the size
of the subspace $k$ increases. The critical size $k_c$ for impure
states at which the state is no longer provably entangled is now
given by
\begin{equation}\label{k-critical-sigma}
k_c=\lceil 2p\frac{\ln{(1+\sigma)}}{\ln{2}}\rceil
\end{equation}
where $\lceil x \rceil$ represents the next higher integer to $x$
(only integral numbers of spins are possible). It is straightforward
to see that with $p$ partially polarized spins, polarization can be
shared over a smaller number of qubits than when the $p$ spins are
ideally polarized, as shown in Figure~\ref{graphics2}.

\section{Conclusions}\label{section-conclude}
In this paper, we have shown that given $p$ perfectly polarized
spins, we can share polarization among $2p-1$ spins, such that their
bias keeps them in the provably entangled region. For example, for
$p=2$ pure spins, we can spread the polarization over $3$ spins,
while ``protecting'' the entanglement.  Similar results can be
achieved when the initial spins are not completely polarized: from
(\ref{k-critical-sigma}), we find that for $\sigma\geq
(\exp{(3\ln{2}/4)}-1) \approx 0.682$, it is possible to spread the
polarization of $2$ qubits onto a higher spin subspace. Our results
suggest that polarization sharing is of limited value with only a
single pair of protons from \pHH\ added onto our substrate molecule
\cite{anwar04}, but show more promise for a higher number of
molecules $M$. Preparing such molecules is in principle possible and
we are currently investigating approaches for such multi-qubit
systems.

\begin{acknowledgments}
We thank the EPSRC for financial support.  MSA thanks the Rhodes
Trust for a Rhodes Scholarship and the NCP, Pakistan for a
post-doctoral fellowship.  We thank Hilary Carteret for helpful
conversations.
\end{acknowledgments}

\appendix*
\section{}
When sharing polarization from $p$ pure qubits, the system always
lies within the regions \textbf{E} and \textbf{ES}, never entering
the explicitly separable region \textbf{S}.  To prove this we must
show that the extracted purity is always greater than the
Gurvits--Barnum bound,
\begin{equation}\label{lower-bound-interscet}
\frac{2^p-1}{2^k-1}>\frac{3}{2(6)^{k/2}};\quad\quad \forall\ k\ge p.
\end{equation}
\begin{proof}
The left hand side of Equation~\ref{lower-bound-interscet} will be a
minimum when $p=1$, and so it suffices to prove this extreme case:
\begin{equation}\label{proofs1}
\frac{1}{2^k-1}>\frac{3}{2(6)^{k/2}}.
\end{equation}
Now the L.H.S. of \eqref{proofs1} is clearly greater than $1/2^k$.
So the inequality will be true when the ratio of $1/2^k$ and right
hand side of \eqref{proofs1} is greater than unity. This ratio
$(2/3)^{1-k/2}$ is greater than one for $k>2$, whereas
\eqref{proofs1} can also be shown to be correct for $k=2$ by
explicit calculation.
\end{proof}

\end{document}